# Factors Affecting QoS in Tanzania Cellular Networks

Adam B. Mtaho and Fredrick R. Ishengoma

*School of Informatics and Virtual Education, The University of Dodoma, Tanzania*

*Abhassigiye@yahoo.co.uk, fishengoma@udom.ac.tz*

**Abstract**

Quality of service (QoS) in cellular communication system is a topic that recently has raised much interest for many researchers. This paper presents the findings obtained from the study on factors affecting QoS in Tanzania cellular networks**.** The study was carried out in Dodoma Municipal, Tanzania. The study employed cross-sectional research design. Information was gathered from structured questionnaire of 240 subscribers during the study of quality of service for the four leading cellular networks in Tanzania (Vodacom, Airtel, Tigo and Zantel). Both qualitative and quantitative data from field survey were collected and analyzed using Statistical Package for Social Sciences (SPSS-version 11) and Excel software. The study findings show that the major factors that degrade QoS in Tanzania cellular networks are inadequate network infrastructure, lack of fairness from service providers and little efforts taken by the government in enforcing the national agreed standards. Other factors are lack of reliable end-to-end systems, geographical terrain, low quality handsets, poor government monitoring on standards and lack of subscriber's skills and training.

**Keywords: Cellular networks, Network coverage, Network capacity, Quality of Service**

## I. Introduction

The cellular communication services in Tanzania were introduced after the liberalization of the communication sector in 1993. Since then, a number of subscribers and vendors have been increasing. Tanzania Communication Regulatory Authority (TCRA) statistics as of September 2013, indicates that the mobile telephone market is the fastest growing sector, with more than 26 million subscribers in a population of about 43 million, of which Vodacom is leading with, about 10 million subscribers, followed by Airtel 8.7 million, Tigo 6.2 million, and Zantel 1.7 million (TCRA, 2013).

Quality of Service (QoS) refers to the effect of service performance that defines the level of a user's satisfaction with the service. The Tanzania Communications QoS Regulations clearly recommends measuring the QoS provided in cellular networks from time to time and comparing them with the norms so as to assess the level of performance. Several factors affect the quality of services in cellular networks. However, so far, very little is known about the QoS in cellular networks in the country. This paper therefore examines factors that affect QoS in Tanzania cellular networks.

The rest of the paper is organized as follows: We start by providing an overview of cellular networks in section 2. Section 3explains the study methodology used. Section 4 presents data collection and data analysis. Study findings are presented in Section 5. We conclude in Section 6 along with recommendation for future practice.

II. **Overview of Cellular Networks**

Cellular communications worldwide has experienced a rapid growth in the past two decades. Cellular phones allow a person to make and receive a call from almost everyplace through a given network. Also, cellular networks enable a person in motion to continue with phone conversation. Cellular communication is supported by an infrastructure called a cellular network, which integrates cellular phones into the Public Switched Telephone Network (PSTN) (Zhang and Stojmenovic, 2005). Apart from voice service, cellular telephony provides a number of services to the users. Such services include Short Message Services (SMS), Instant Messaging (IM), and Multimedia Messaging (MMS). SMS allows a subscriber to send and receive text message using the mobile terminal. IM is a short message exchanged between mobile users in real time while MMS allows users to send pictures, video or voice to other mobile terminals or an electronic mail account. Other services include e-mail services, wireless Internet, emergence calls, video services and mobile TV (Pashtan, 2006). A cellular network provides cell phones or mobile stations (MSs), with wireless access to PSTN (Zhang and Stojmenovic, 2005)

*A. Essential Elements of a Cellular Networks*

Mobile Stations(MS) essentially enables a user uses to make communication. The MS is made up of two parts, the handset and Subscriber Identity Module (SIM), whose major function is to store data for both the operator and subscriber (Mishra, 2004). The service coverage area of a cellular network is divided into several smaller areas, referred to as cells, each of which is served by a base station (BS or BTS). The BS is fixed, and it is connected to the Mobile Telephone Switching Office (MTSO), also known as the Mobile Switching Center (MSC). An MTSO is in charge of a cluster of BSs and it is, in turn, connected to the PSTN (Zhang and Stojmenovic, 2005). For GSM network, BTS is then connected to MSC via the Base Station Controller (BSC). The BSC main function is usually to handle radio resource management and handovers of the calls from one BTS (or cell/sector) to another BTS equipped in it (Mishra, 2004). With the wireless link between the BTS and MS, MSs are able to communicate with wire line phones in the PSTN. Both BSs and MSs are equipped with a transceiver. Figure 1 illustrates a typical cellular network, in which a cell is represented by a hexagon and a BS is represented by triangle.

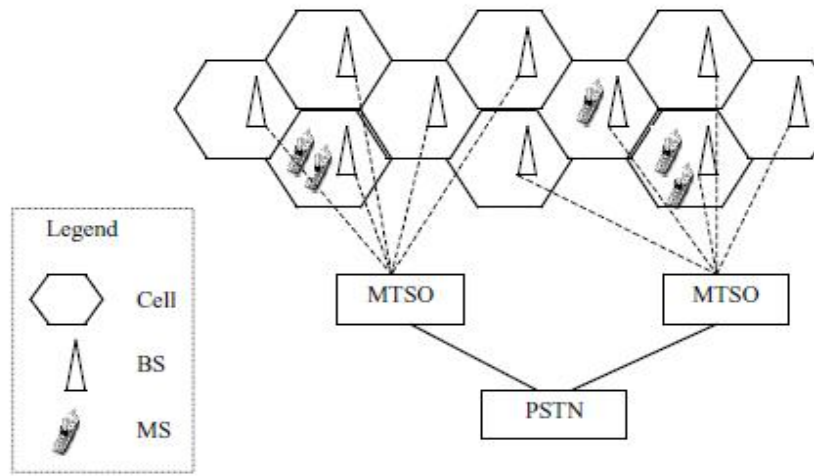

**Figure. 1.** Typical cellular network *(Source: Zhang, J. and Stojmenovic, I., 2005)*

A cellular network is a network that divides a geographic area into cells such that the same radio frequency (RF) can be reused in two cells that are a certain distance apart (Zhang and Stojmenovic, 2005). The cellular network operates based on the frequency reuse concept, which offers very high capacity in a limited spectrum allocation. Each BTS is allocated a segment of the total number of channels available to the entire system, and nearby BTS are assigned non-overlapping channel sets. All available channels are later assigned to a relatively small number of neighboring BTSs. The BTS is then used within a small geographic area called a cell. In this way these few available channels are distributed throughout the geographic region and may be reused as many times as necessary, hence providing effective utilization of frequency while providing adequate network coverage within the area (Rappaport, 2002).

When the demand for service increases due to increase of subscribers, the number of BTSs may be increased, so as to increase the corresponding number of channels. This provides additional radio. Some of the features that determine cellular network performance are network coverage and capacity. Both capacity and coverage have direct impact on QoS. Network coverage refers to the maximum allowed distance from which the MS can be reliably connected to the nearby BTS (Mishra, 2004). Network capacity on other hand refers to the maximum number of subscribers a given BTS can cater at a given time. The higher the capacity, the more the number of subscribers who can be allowed to access cellular service at a given time (Mishra, 2004).

III. **Study Methodology**

   A. *Study Location and Justification for Its Selection*

The study was conducted in Dodoma Municipal, Tanzania. Dodoma Municipal was selected because it is a rapidly growing town that needs reliable cellular communication services for its sustainable growth. Moreover in this municipality all four leading cellular networks in Tanzania (Airtel, Tigo, Vodacom and Zantel) are present. The study employed the cross-sectional research design, whereby data was collected among six institutions in Dodoma Municipal. Information from structured questionnaire of 240 students' subscribers, six key informants and observation was collected to assess the quality of network performance and online customer service.

### B. Sampling Procedures
#### a. Population

The population from which the sample was drawn for this study involved subscribers who are getting cellular communication services from the four cellular networks present in Dodoma (Tigo, Vodacom, Zantel and Airtel). The sampling frame was a students' subscriber from six higher learning institutions namely: College of Business Education (CBE), Institute of Rural Development Planning (IRDP), The University of Dodoma – College of Informatics and Virtual Education (CIVE), The University of Dodoma – College of Health and Social Science (CHSS), The University of Dodoma – College of Education (CoED) and St.John's University of Tanzania (SJUT) in Dodoma Municipal.

#### b. Sample Size and Sampling Technique

According to the Department of Research and Consultancy of the Dodoma Municipal, Dodoma Municipal has the total population of 507,141 people. In order to get the sample size the researcher adopted the formula that has been developed by Yamena (1967).

$$n = \frac{N}{(1 + n(e)2)}$$

Where $n$ – Sample Size, $N$ – Total Population, $e$ – Detection error expressed into percentage $(5\% - 10\%)$, and thus from $N = 507,141, e = 6.74\%$,

$$n = \frac{507,141}{\left(1 + 507,141 \left(\frac{6.74}{100}\right)2\right)}$$

$$n = 220 \text{ Respondents}$$

That means, according to Yamena (1967) the respondents were not required to be less than 220. In order to make the research manageable, the sample size of 240 subscribers from six higher learning institutions was involved in this study. The 240 respondents were selected by employing purposive sampling technique. The specific criterion for selecting respondents was; being student subscriber who lives and study at the given institution.

Later, 40 respondents were randomly selected from each educational institution using simple random sampling procedure to make a total of 240 respondents. Furthermore, six key informants (two site engineers, one frequency engineer and three cellular network engineers) were interviewed using a checklist. Selection of the key informants was based on the experience in working in cellular networks industry. It was logically assumed that the sample represented the whole population of mobile subscribers in Tanzania.

IV. **Data Collection and Data Analysis**

In this study, for primary data, structured questionnaires, in-depth interview, and observation were used as primary data collection instruments while documentary review were used for the collection of the secondary data. Both qualitative and quantitative data from field survey were collected and analyzed using Statistical Package for Social Sciences (SPSS-version 11) and Excel software program. Data were first coded in the form suitable for addressing research questions and the method of analysis to be employed. The analysis of the recorded and summarized data from key informants used ethnographic approach; that is relying on the direct information given by respondents according to the check list used during the discussion.

V. **Study Findings**

A. *Factors that Affects the QoS in Tanzania Cellular Networks*

Several factors that affect QoS in Tanzania cellular networks were identified during the study. Among the factors pointed out to affect QoS by customers include limited coverage and capacity, poor government monitoring on standards and lack of skills and training on the use of mobile phones. Other factors include lack of fairness from service providers, low quality handsets and delay in allocation of adequate network infrastructure. Further, geographical terrain and lack of reliable end-to-end systems were pointed by telecommunication engineering professionals to affect QoS in cellular networks.Figure 2 below presents the size and percentage response of customers on the factors that affect QoS in Tanzania cellular networks.

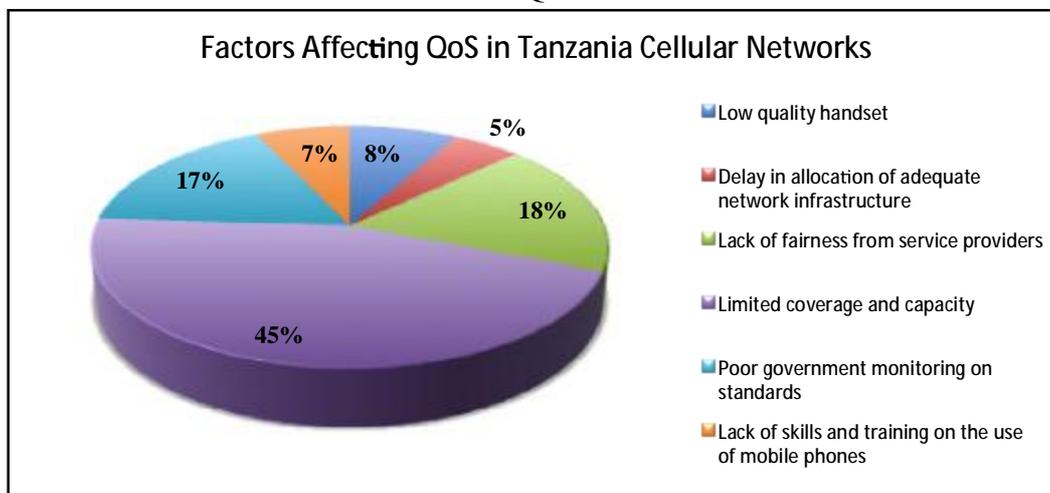

**Figure. 2.**Response of customers on the factors affecting QoS in Tanzania cellular networks.

Figure 2 shows a number of factors that degrade the QoS in Tanzania cellular networks. These factors are discussed as follows:

a. *Limited Network Coverage and Capacity*

Lack of adequate network coverage and capacity are the leading causes that affect QoS in Tanzania cellular networks as it was evidenced by 45% of the total responses. The study found that network infrastructure is available mainly in urban areas and limited in rural areas. According to Tanzania's Infrastructure report (2010), only 65 percent of the Tanzania population lives within the range of a GSM signal, compared with over 90 percent in neighboring Kenya and Uganda. This indicates that network coverage in Tanzania is still inadequate.

The study found that despite having moderate network coverage in urban areas, the network capacity is still limited. For example, despite having more than 23, 000 subscribers at UDOM area has only three BTS. This makes an average of about 7500 subscribers per single BTS. This number is too large for one BTS. It does nor matter which capacity improvement techniques is used. The study found that network related problems (such as network outage and call dropout) are among the reasons why subscribers own more than one SIM cards. According to the findings, 44.6% of subscriber complained about network related problems to be the main causes for subscribers owning more than one SIM card from different operators. Since QoS in cellular networks depends mainly on network coverage and capacity (Mishra, 2004), inadequate network coverage and low network capacity contributes to degradation of QoS in Tanzania cellular networks.

### b. *Lack of Skills and Training*

Lack of skills and training on using mobile phones is another cause that affects QoS as it was evidenced by 7% of the total responses. Using mobile phone requires knowledge and ability to properly follow the procedures. For example, the study found that some of subscribers do not charge the phones to their full capacity. Improper charging of the battery limits its life span. Lack of such knowledge can compromise the quality of the handset. The handset can be easily susceptible to call drop out and network outage problems if its battery strength is weak. The study found that some of the subscribers do not know how to effectively use mobile phones. Most of the mobile handsets are configured to operate in English or other foreign languages, while majority of Tanzanians understand Kiswahili. Respondents pointed out this problem as the source of difficulty in using mobile phone. Improper knowledge of how to buy good quality phones was also reported to be the major bottleneck towards aching effective QoS. This is because poor quality handsets suffer from voice quality degradation (Mishra, 2007)

### c. *Poor Government Monitoring on Standards*

Poor government monitoring on standards is another cause that affects QoS as it was evidenced by 17% of the total responses. The study found that despite there being national standards, operators are following their own standards. For instance, a caller in Airtel network may get informed about the time and amount of money spent instantly after finishing the call. Vodacom and Tigo networks do not give such information. This implies that Airtel subscribers can easily verify the bills of their calls. The study found that operators use the same standards for few services such as using the same dialing code to request online customer care (code 100 is used), air balance inquiry and air time recharging.

### d. *Lack of Fairness from Service Providers*

Lack of fairness from service providers is reported to be another barrier towards achieving effective QoS in cellular networks. 18% of the total responses pointed out that providers were sometimes unfair to the customers. For example, the study found that if a call drops out due to network problems, the service provider does not pay the user for the loss incurred.

The International Telecommunication Union (ITU) recommends that to ensure service level quality, service level agreement (SLA) must be included in contractual agreements between customers and service providers. This includes tariffs and billing, network performance, and compensation for unachieved level of service quality. The study found that SLA is not practiced in Tanzania. For example, registration in cellular networks service does not include SLA between mobile subscriber and service providers. In other words, the customer is neither guaranteed by the provider to get high quality service nor assured to be compensated if the normal service quality level is not achieved. Lack of SLA between the operator and subscriber therefore affects QoS.

Lack of clear information about the service provided by network operators is another example that shows how operators are unfair to their customers. For example the study found that subscribers are always told to subscribe to the caller tune service. However, subscribers are not informed how to unsubscribe from this service. As reported by Materu and Dyamett (2010), subscribers are instructed to subscribe i.e. to join with new services, but they are not informed or taught how to unsubscribe. According to Tanzania Communications (QoS) Regulations, 2005, service providers must ensure that customers are provided with information about the services that they provide, so as to enable them to make decision. However, the study found that such rule is not effectively followed by operators.

e. *Low Quality Handsets*

In response to the factors affecting the QoS in cellular networks, the lack of handsets with good quality was one of the complaints. This contributed 8% of the total responses. Low quality handsets affect the QoS. Usually handsets that receive low network signal uses more power to remain connected, than the handset near the BTS (Mishra, 2004). According to Ditech Networks (2011), in CDMA networks, the use of inexpensive low quality handsets allows hybrid and acoustic echoes; resulting degraded service quality. It is therefore recommended to have mobile phone with good quality.

f. *Delay in Allocation of Adequate Network Infrastructure*

Delay in allocation of BTSs is another problem. It contributes 5% of the total responses. The respondents pointed out that operators have been delaying to introduce new BTSs despite the rapid increase of customers in the area. It was pointed out by the respondents that the increase of subscribers does not match with the increase of BTSs. For example, the study found that there are only three BTSs serving the population of about of 23000 subscribers at UDOM. This makes an average of 7500 subscribers per BTS. This number is too large to be accommodated by a single BTS. The study found that operators delay to increase number of BTS regardless of the rising demand due to high investment and running cost of the equipment.

g. *Geographical Terrain*

Geographical terrain was pointed by cellular network engineers to be one of the variables that affect the QoS. It was stated that around Dodoma Municipal there are hills and mountains. Radio waves require direct line of sight from BTS to BTS. Presence of hills and mountains create obstacles and prevents signal of one BTS from reaching another BTS or BSC. This problem degrades QoS and affects both network coverage, and capacity. Presence of hills and mountains in Dodoma municipals were pointed out by cellular network engineers to bring difficulties during network planning. For example, the study found the Vodacom BTS at UDOM, which is positioned about 9 kilometers from the BSC is linked with the BSC via another BTS at Kisasa (about 12 Km from UDOM). Redirection of UDOM BTS was done to avoid unfavorable signal propagation caused by obstacles due to the hilly terrain along the UDOM area.

### h. *Lack of Reliable End to End Systems*

Relaying on radio link as the major connection between BSC and MSC was pointed out by site and frequency engineers to be another factor that affects QoS in cellular networks. Despite the introduction of fiber optic cable by SEACOM in July 2009, cellular networks in Tanzania still use radio link as the primary media for signal transmission from BSC to MSC. Radio link suffers from limited throughput and electromagnetic interferences from other sources. Interference cause noise, call drop and call set up failure. However, the study found that Vodacom is cellular network operator in Tanzania who currently uses fiber optic to link BSC and MSC as an alternate to radio link.

## VI. Conclusions

In this paper, we presented the factors that affect the Quality of Service in Tanzania cellular networks. Some factors that were identified to be the contributors of poor QoS in Tanzania cellular networks includes; limited network coverage and capacity, inadequate network infrastructure, lack of fairness from service providers and little efforts taken by the government in enforcing the national agreed standards. The study recommends TCRA to ensure that the established standards are met and followed by operators. Further, telecomm operators in the country are advised to improve network coverage and provide reliable and fair services to customers.